\documentclass[letterpaper,twocolumn,english,aps,prb,final,superscriptaddress,showpacs]{revtex4}
\usepackage[T1]{fontenc}
\usepackage[latin9]{inputenc}
\usepackage{amsmath}
\usepackage{amssymb}
\usepackage{graphicx}

\makeatletter

\pdfpageheight\paperheight
\pdfpagewidth\paperwidth

\providecommand{\tabularnewline}{\\}

\@ifundefined{textcolor}{}
{%
 \definecolor{BLACK}{gray}{0}
 \definecolor{WHITE}{gray}{1}
 \definecolor{RED}{rgb}{1,0,0}
 \definecolor{GREEN}{rgb}{0,1,0}
 \definecolor{BLUE}{rgb}{0,0,1}
 \definecolor{CYAN}{cmyk}{1,0,0,0}
 \definecolor{MAGENTA}{cmyk}{0,1,0,0}
 \definecolor{YELLOW}{cmyk}{0,0,1,0}
}

\usepackage{babel}
\usepackage{babel}
\usepackage{babel}
\usepackage{times}\usepackage{xspace}\usepackage{times}\usepackage{amsfonts}\usepackage{babel}

\usepackage{dcolumn}

\renewcommand{\mathbf}[1]{\ensuremath{\boldsymbol{#1}}}\allowdisplaybreaks

\makeatother

\usepackage{babel}
\begin{document}

\title{Electric dipole sheets in BaTiO$_{3}$/BaZrO$_{3}$ superlattices}

\author{Zhijun Jiang }

\affiliation{Electronic Materials Research Laboratory--Key Laboratory of the Ministry
of Education and International Center for Dielectric Research, Xi'an
Jiaotong University, Xi'an 710049, China}

\author{Bin Xu}

\affiliation{Physics Department and Institute for Nanoscience and Engineering,
University of Arkansas, Fayetteville, Arkansas 72701, USA }

\author{Fei Li}

\affiliation{Electronic Materials Research Laboratory--Key Laboratory of the Ministry
of Education and International Center for Dielectric Research, Xi'an
Jiaotong University, Xi'an 710049, China}

\author{Dawei Wang}

\email{dawei.wang@mail.xjtu.edu.cn}

\affiliation{Electronic Materials Research Laboratory--Key Laboratory of the Ministry
of Education and International Center for Dielectric Research, Xi'an
Jiaotong University, Xi'an 710049, China}

\author{C.-L. Jia}

\affiliation{Electronic Materials Research Laboratory--Key Laboratory of the Ministry
of Education and International Center for Dielectric Research, Xi'an
Jiaotong University, Xi'an 710049, China}

\affiliation{\textsuperscript{}Peter Grünberg Institute and Ernst Ruska Centre
for Microscopy and Spectroscopy with Electrons, Research Center Jülich,
D-52425 Jülich, Germany}
\begin{abstract}
We investigate two-dimensional electric dipole sheets in the superlattice
made of BaTiO$_{3}$ and BaZrO$_{3}$ using first-principles-based
Monte-Carlo simulations and density functional calculations. Electric
dipole domains and complex patterns are observed and complex dipole
structures with various symmetries (e.g. $Pma2$, $Cmcm$ and $Pmc2_{1}$)
are further confirmed by density functional calculations, which are
found to be almost degenerate in energy with the ferroelectric ground
state of the $Amm2$ symmetry, therefore strongly resembling magnetic
sheets. More complex dipole patterns, including vortices and anti-vortices,
are also observed, which may constitute the intermediate states that
overcome the high energy barrier of different polarization orientations
previously predicted by Lebedev [\onlinecite{Lebedev2013}]. We also
show that such system possesses large electrostrictive effects that
may be technologically important. 
\end{abstract}


\maketitle

\section{Introduction}

Two-dimensional (2D) magnetic systems have been an important topic
in the study of magnetism \cite{Manousakis1991,Stamm1998,Steele2011}.
The Ising model is such an important theoretical model that it had
been investigated using analytical (e.g., mean-field theory) and numerical
{[}e.g., Monte-Carlo (MC) simulations{]} methods \cite{Morgenstern1981,Gawlinski1985,Moessner2001}.
More than just theoretical interest, certain real magnetic systems
indeed have special magnetic structures, in particular the so-called
magnetic sheets, i.e., layers containing magnetic moments separated
by non-magnetic layers \cite{Coey}, examples including K$_{2}$NiF$_{4}$
and Sr$_{2}$RuO$_{4}$ \cite{Matzdorf2000}. Such systems have interesting
properties (e.g. nanoscale magnetic domains) and a common feature
of them is that the energy difference between different structure
and magnetic phases is very small \cite{Matzdorf2000,Iglesias2002,Jagla2004}.
On the other hand, it is interesting to note that similar structures
comprised of \emph{electric} dipoles are much less known. The concept
of ``arrays of nearly independent ferroelectrically ordered quasi-two-dimensional
layers'' was first clearly proposed by Lebedev in his investigation
of KNbO$_{3}$/KTaO$_{3}$ superlattices (SLs) \cite{Lebedev2011}.
Here, via first-principles-based MC simulations, we will further point
out the analogy between electric dipole sheets (quasi-two-dimensional
ferroelectric layers) existing in the SL made of BaTiO$_{3}$ (BTO)
and BaZrO$_{3}$ (BZO) to magnetic dipole sheets in terms of energy
degeneracy and the complexity of dipole patterns.

The solid solution of BTO and BZO, which is chemically disordered
and denoted by Ba(Zr,Ti)O$_{3}$ (BZT), is a well known ferroelectric
material, most famous for its relaxor properties (see, e.g., Refs.\cite{Maiti2008,Shvartsman2009,Maiti2011,Kleemann2013,Akbarzadeh2012,Prosandeev2013,Prosandeev_2013,Sherrington2013,Pirc2014,Wang2014}
and references therein). Among its anomalous properties, the best
known signature is perhaps the shift of the temperature $T_{m}$,
where the susceptibility as a function of temperature peaks, with
frequency used in the measurement of the susceptibility. Recently,
Akbarzadeh \textit{et al.} \cite{Akbarzadeh2012} developed a first-principles-based
effective Hamiltonian to investigate disordered BZT. Based on this
effective Hamiltonian, Sherrington \cite{Sherrington2013} argued
that BZT can be understood as soft spin-glass. Using the same effective
Hamiltonian, MC simulations have shown that dipoles on Ti-containing
sites are substantially larger than those on Zr-containing sites,
which provides the possibility of having isolated electric dipoles
in this type of materials. For instance, a SL made of alternating
layers containing Ti and Zr ions will give rise to isolated planes
of electric dipoles that resembles magnetic sheets.

However, it is unclear if such 2D electric dipole system indeed behaves
like magnetic sheets in the sense that very complex dipole patterns
exist. Note the recent study indicate large band of phonon frequencies
along $\Gamma$-$Z$-$R$-$X$-$\Gamma$ have similar large negative
values, indicating various instabilities, but only certain high-symmetry
dipole patterns are computed via \emph{ab initio} computations and
compared to the ferroelectric ground state\cite{Lebedev2013}. Moreover,
on the application side, experiments have found that electrostriction
effect is generally stronger in ordered systems than in disordered
systems \cite{George and Bellaiche2001,Li2014}. It is important to
know if such effect also happens with lead-free BZT. As a matter of
fact, disordered BZT contains Ti ion under various chemical environment,
for instance, there could be 1, 2, $...$, $6$ Zr ions surrounding
one Ti ion as its nearest neighbors. In addition, the distance from
other Ti ions to a given Ti ion is also random. This randomness potentially
harms physical properties of BZT that require the correlation of dipoles
on many Ti ions. Ordered BZT offers certain level of freedom to address
these two problems together because: (i) the distance between Ti ions
are preconditioned; (ii) the chemical environment are uniform for
every electric dipole in the 1:1 SL. In addition, confined dimensionality
may also be an advantage to certain applications \cite{Schlom2013}.

Therefore investigating chemically ordered BZT will expand our understanding
of lead-free relaxors and may suggest systems that have better physical
properties. Here we consider chemically ordered BZT SLs with alternating
Ti and Zr layers grown along the $\left[001\right]$ direction. In
such a structure, unlike the disordered BZT, large electric dipoles
are restricted to each Ti-containing $\left(001\right)$ planes \cite{Akbarzadeh2012},
which significantly modifies the energy landscape of the system. As
a consequence, novel phases are allowed and complex electric dipole
patterns may form. In addition, the chemical environment of each large
electric dipole is the same (since every Ti ion is surrounded by 4
Ti and 2 Zr ions as nearest neighbors), therefore giving rise to a
single Debye-type relaxation mode in its dielectric responses, and
distinguishing itself from disordered BZT. Moreover, comparing the
electrostriction of ordered BZT to disordered BZT systems, we find
that ordered BZTs indeed possess large electrostrictive coefficient,
which may be exploited in applications. We also note such a structure
is one of the ferroelectric-dielectric SLs,\textbf{ }structurally
similar to the BaTiO$_{3}$/ SrTiO$_{3}$, PbTiO$_{3}$/ SrTiO$_{3}$
and Pb(Zr$_{0.3}$,Ti$_{0.7}$)O$_{3}$/ SrTiO$_{3}$ SLs, which are
actively investigated theoretically and experimentally\cite{Lisenkov2007,Li2007,QZhang2013,ibm2014}.

\section{Method}

Practically, we simulate BZT SLs with alternating Ti and Zr layers
along the $\left[001\right]$ direction {[}shown in the inset of Fig.
\ref{fig:The-dipole-configuration}(a){]}. The first-principles-based
effective Hamiltonian developed in Ref. [\onlinecite{Akbarzadeh2012}]
is employed. In this effective Hamiltonian, the internal energy is
given by $E_{\textrm{int}}(\{\mathbf{u}_{i}\},\{\mathbf{v}_{i}\},\{\eta_{H}\},\{\sigma_{j}\})$.
Here $\mathbf{u}_{i}$ is the local (Zr or Ti-centered) soft mode
in unit cell $i$, which is directly proportional to the local electric
dipole moment in that cell. $\{\mathbf{v}_{i}\}$ is the Ba-centered
dimensionless local displacements that are related to the inhomogeneous
strain variables inside each cell \cite{Zhong1994,Zhong1995}, while
$\eta_{H}$ is the homogeneous strain tensor. Finally, $\{\sigma_{j}\}$
characterizes the atomic configuration of the BZT solid solution,
that is $\sigma_{j}=$ +1 or -1 corresponds to the presence of a Zr
or Ti atom located at the lattice site $j$, respectively. 

The effective Hamiltonian for perovskites approach was first developed
in Ref. [\onlinecite{Zhong1995}], which contains five energy terms:
(i) the local mode self-energy; (ii) the long-range dipole-dipole
interaction; (iii) the energy due to short-range interaction; (iv)
the elastic energy; and (v) the energy due to the interaction between
local mode and strain. While the first two terms are directly related
to dipoles, which is important for ferroelectric materials, the last
three terms are all related to elastic energy. It is worth noting
that the effects of $\mathbf{u}_{i}$ are two-fold: (i) it represents
the local mode (see Eq. (5) of Ref. [\onlinecite{Zhong1995}]); (ii)
it represents the local lattice distortion (since its origin is an
optic phonon), and thus giving rise to an elastic energy, namely the
short-range interactions, as shown in Eq. (9) of Ref. [\onlinecite{Zhong1995}].
In addition to the short-range interaction, the effective Hamiltonian
also includes both homogeneous strain ($\eta_{H}$) and inhomogeneous
strain ($\mathbf{v}_{i}$). Therefore elastic energies have been properly
included in our effective-Hamiltonian-based MC simulations.

In this study, the distribution of the Zr and Ti ions is chosen to
mimic the SL structure, which are kept frozen in all the simulations.
We use a $12\times12\times12$ supercell (8640 atoms) to mimic such
a system and 160,000 MC sweeps to relax the system at each temperatures.
The system is cooled down from 1000\,K to 10\,K with a step size
of 10\,K to obtain static properties. In addition, direct \emph{ab
initio} calculations are performed to verify the novel phases observed
in MC simulations \cite{no_AFD}.

\section{Results and discussion}

\begin{figure}
\includegraphics[width=8cm]{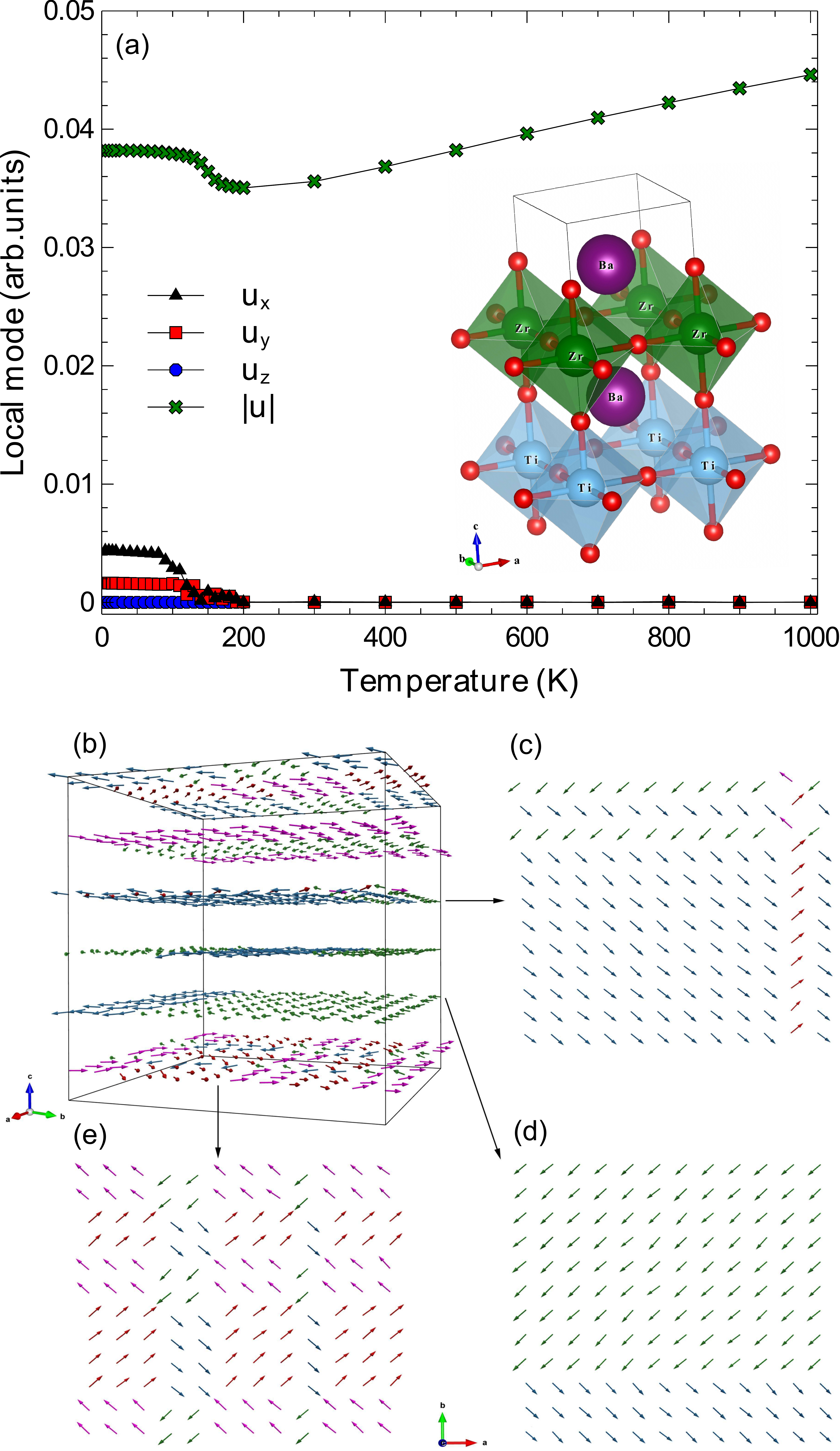}

\protect\caption{(Color online) The average Cartesian components and absolute magnitude
of the local mode vectors as a function of temperature are shown in
Panel (a). The inset of Panel (a) displays the unit cell of the BZT
SLs. A snapshot of dipole configuration of the BZT SLs at 10\,K is
shown in Panel (b). The dipole structures from the first, second,
and fourth layer (counting from bottom along $z$ and only layers
containing Ti ions are counted) are shown in Panels (c)--(e). Four
different colors are used to denote dipoles of different orientations
(one of the four $\left\langle 110\right\rangle $ directions). \label{fig:The-dipole-configuration}\label{fig:phase_transition}}

\end{figure}

Let us first examine the phases of the BZT SLs at different temperatures.
Figure \ref{fig:The-dipole-configuration}(a) shows the supercell-averaged
local mode $\left\langle u_{x}\right\rangle $, $\left\langle u_{y}\right\rangle $,
$\left\langle u_{z}\right\rangle $ and the average of magnitude,
$\left\langle \left|\mathbf{u}\right|\right\rangle $. As we can see,
$\left\langle \left|\mathbf{u}\right|\right\rangle $ is always larger
than $\left\langle u_{i}\right\rangle $ ($i=x,y,z$), indicating
that dipole orientations still have strong randomness even at the
lowest temperature. In other words, this system remain macroscopically
paraelectric down to the lowest temperatures although $\left\langle u_{x}\right\rangle $
and $\left\langle u_{y}\right\rangle $ have slightly larger magnitudes
below $T\simeq100$ K, indicating a phase transition happens at $T_{C}\sim150$\,K
(note $\left\langle u_{z}\right\rangle $ is zero for all temperatures).
To understand the temperature dependency of local modes, Fig. \ref{fig:The-dipole-configuration}(b)
shows a snapshot of the dipole configuration of the system at 10\,K,
which exhibits some interesting features: (i) Large dipoles only exist
on Ti sites, therefore only 6 layers of dipoles can be seen; (ii)
All the dipoles stay inside the $\left(001\right)$ planes, pointing
along one of the four $\left\langle 110\right\rangle $ directions;
(iii) In a given Ti layer, most of the dipoles have the same $y$
component $d_{y}$, but their $x$ component, $d_{x}$, can be different
and form 90$^{\circ}$ domains {[}see Figs. \ref{fig:The-dipole-configuration}(c)
and \ref{fig:The-dipole-configuration}(d){]}; (iv) Two dipoles from
different Ti layers can be parallel, anti-parallel, or perpendicular
to each other, indicating the correlation between dipoles in different
layers is weaker than that within the same layer. Interestingly, these
observations suggest a system that strongly resembles magnetic sheets,
where magnetic planes are separated by nonmagnetic materials. For
instance, in K$_{2}$NiF$_{4}$, the antiferromagnetic Ni$^{2+}$
does not couple between adjacent planes \cite{Coey}. Here in BZT
SLs, magnetic dipoles are replaced by electric dipoles. Moreover,
it appears that one rule of forming dipole patterns is to avoid head-to-head
and tail-to-tail dipole pairs. But exceptions do exist as Figs. \ref{fig:The-dipole-configuration}(c)
and \ref{fig:The-dipole-configuration}(e) show the existence of vortices
and anti-vortices, which form at the intersections of four nanodomains
(shown in different colors) and contain head-to-head or tail-to-tail
dipole pairs. Such vortex and anti-vortex structures are known in
magnetic systems \cite{Lee1986} and BiFeO$_{3}$ \cite{Balke2012}.
The existence such structures in the BZT SLs further emphasizes the
similarity between electric dipole sheets and magnetic dipole sheets.

Clearly, the two observations (iii) and (iv) explain why $\left\langle u_{x}\right\rangle $
and $\left\langle u_{y}\right\rangle $ are minuscule even at the
lowest temperature. In addition, similar to the magnetic sheets, where
2D Heisenberg model had shown various interesting phase transitions
and patterns of magnetic moments \cite{Coey,Sereba1993,Siurakshina2001,Blote2002,Deng2002},
Fig. \ref{fig:The-dipole-configuration} shows that the 2D electric
dipole system can also have complicated patterns, resembling a frustrated
system. The observations (i)--(iv) can be easily explained qualitatively.
First, since the lattice constant of BZO is larger than BTO \cite{BZT-lattice-constant},
Ti layers effectively experience a tensile strain in the $x$-$y$
plane and it is hard to develop local mode along $z$ because in this
direction the lattice is compressed. Second, it is well known that
at low temperature BTO has the $R3m$ phase with local dipoles along
$\left[111\right]$ \cite{Zhong1994}. Here since polarization along
the $\left[001\right]$ is disfavored, it is natural that the dipoles
point along $\left\langle 110\right\rangle $. Third, since Zr has
a larger radius, when BTO and BZO are put together (as in the SL),
effectively, the BTO layers has a larger lattice constant, and therefore
smaller short-range interaction \cite{Zhong1995}, than its natural
state, therefore the ions may have more freedom in their movement
(i.e., not necessarily along a singe $\left\langle 110\right\rangle $
direction). In addition, the dipole-dipole interaction is also weakened
since the dipole-dipole interaction is reduced from 3D to 2D. This
explains why in the Ti layers, the orientations of dipoles seem to
have more freedom. Finally, the dipole-dipole interaction between
Ti layers is weak as the distance is doubled compared to that of BTO.
In addition, the interaction could be screened due to the existence
of Zr layers. This explains why there is essentially no dipole correlations
between Ti layers.

\subsection{Various dipole configurations}

\begin{figure}
\includegraphics[width=9cm]{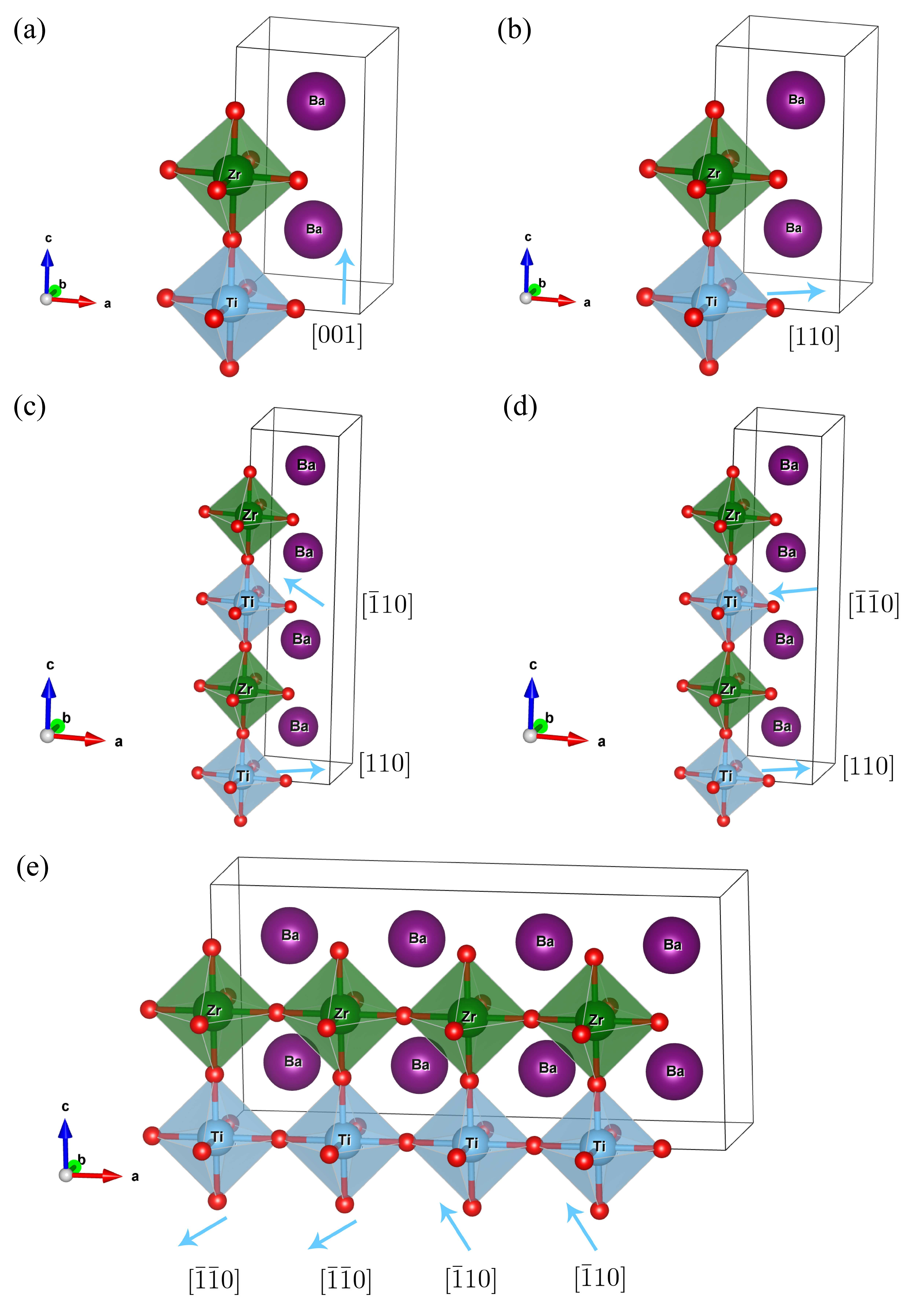}

\protect\caption{(Color online) Five dipole configurations are shown and their total
energies are calculated using density functional theory (see text).
In Panels (a)--(e), the directions of dipoles on each Ti site are
shown as blue arrows (also see Tab. \ref{tab:energy_comparision}).
Four of these configurations in Panels (b)--(e), which are suggested
by MC simulations, are found to be low-energy states. \label{fig:different_phase}}

\end{figure}

The qualitative explanation has obviously invoked arguments that are
rather speculative. To quantitatively understand the complex dipole
configurations in our MC results, we performed \emph{ab initio} computation
using the \textsc{Quantum Espresso} software package \cite{Espresso}.
Two versions of the exchange-correlation functionals {[}namely, the
local density approximation (LDA) \cite{Ceperley_Alder1980} and the
generalized gradient approximation (GGA) of Perdew-Burke-Ernzerhof
(PBE) \cite{PBE1996}{]} are used along with ultrasoft pseudopotentials
\cite{D. Vanderbilt1990} implemented in the GBRV package \cite{GBRV},
in which the Ba \textit{5s 5p 5d 6s}, Zr\textit{ 4s 4p 4d 5s}, Ti
\textit{3s 3p 4s 3d}, and O \textit{2s 2p} orbitals were treated as
valence orbitals. A plane wave basis with kinetic energy cutoff of
1088\,eV was used to ensure the convergence in all the calculations.
$k$-point sampling of $8\times8\times4$ was used for all the 10-atom
supercell, $8\times8\times2$ for the $20$-atom supercell, and $2\times8\times4$
for the $40$-atom supercell (see Fig. \ref{fig:different_phase}).
The initial positions of the atoms are constructed from our MC simulation
results, which are then further relaxed until all the force components
on each atom are less than $10^{-4}$ Hatree/Bohr. We note, in all
the \emph{ab initio} computations except the $P4mm$ structure {[}Fig.
\ref{fig:different_phase}(a){]}, we also find the shift of Zr ions
is much smaller than that of Ti ions, consistent with MC results that
the dipoles are essentially zero on Zr cells. All these results indicate
that our MC results indeed show the correct dipole configurations. 

In our direct \emph{ab initio} computation, we have set up five SLs
with different dipole configurations to compare their energies. Figure
\ref{fig:different_phase}(a) shows a 10-atom cell that has a local
dipole along $\left[001\right]$, which has the $P4mm$ symmetry.
Figure \ref{fig:different_phase}(b) is also a 10-atom cell, but having
in-plane dipoles along $\left[110\right]$ and the $Amm2$ symmetry.
Figures \ref{fig:different_phase}(c) and \ref{fig:different_phase}(d)
show more complicated structures with 20-atom supercells, which mimic
the dipoles in different Ti layers. In Fig. \ref{fig:different_phase}(c),
the dipoles in the two Ti layers are perpendicular to each other (space
group $Pma2$), while in Fig. \ref{fig:different_phase}(d), the two
dipoles are antiparallel to each other (space group $Cmcm$). Both
of these two situations can be found in Fig. \ref{fig:The-dipole-configuration}(b).
Finally, Fig. \ref{fig:different_phase}(e) shows a 40-atom supercell,
which is a simplified version of the complex dipole patterns observed
in a single Ti layer. In this configuration, the first two dipoles
along $\hat{x}$ point along $\left[\bar{1}\bar{1}0\right]$, while
the next two dipoles point along $\left[\bar{1}10\right]$, which
has the $Pmc2_{1}$ symmetry \cite{Yang2012,Yangyurong2014}. In other
words, while every layer has the same dipole pattern, in each layer,
the dipoles are similarly oriented along $\hat{x}$, but their $\hat{y}$
components alternate every two Ti ions along the $\left[100\right]$
direction. It is surprising that, this phase is structurally very
different from that of the $Amm2$ phase (for instance, it has a macroscopic
polarization along $\left[100\right]$, but $Amm2$ has a polarization
along $\left[110\right]$). Note this configuration features dipole
heads connected to other dipoles' tails. It is found that head-to-head
dipole configurations are general not stable in \emph{ab initio} computation
of small supercells shown in Fig. \ref{fig:The-dipole-configuration}
\cite{head-to-head}. We also note that in MC simulations using smaller
supercell, which contains 20 or 40 atoms, all dipole configurations
shown in Fig. \ref{fig:different_phase} {[}except Fig. \ref{fig:different_phase}(a){]}
are also observed, but appear in different cooling processes or at
different MC steps, indicating that these dipole patterns can dynamically
transform from one to another. 

\begin{table*}
\begin{tabular}{|c|c|c|c|c|c|c|}
\hline 
Space group  & $\Delta E$ (LDA) %
\footnote{The energy unit is meV per 10 atoms.%
} & $\Delta E$ (PBE)  & $\Delta E$ (MC)%
\footnote{MC simulations are performed at $T=5$\,K%
}  & Dipole-dipole (MC) & Short-range (MC) & Dipole\tabularnewline
\hline 
$P4mm$ (\#99)  & 88.22  & 179.79  & -- & -- & -- & $\left[001\right]$\tabularnewline
\hline 
$Amm2$ (\#38)  & 0.00  & 0.00  & 0.00 & 0.00 & 0.00 & $\left[110\right]$\tabularnewline
\hline 
$Pma2$ (\#28)  & 2.48  & 3.93  & 0.17 & -0.03 & 0.63 & $\left[110\right]$\&$\left[\bar{1}10\right]$\tabularnewline
\hline 
$Cmcm$ (\#63)  & 2.93  & 3.58  & 0.05 & -0.75 & 1.52 & $\left[110\right]$\&$\left[\bar{1}\bar{1}0\right]$\tabularnewline
\hline 
$Pmc2_{1}$ (\#26)  & 2.05  & 2.96  & 1.37 & -19.01 & 22.56 & $\left[\bar{1}\bar{1}0\right]$\&$\left[\bar{1}10\right]$\tabularnewline
\hline 
\end{tabular}

\protect\caption{The energies of various dipole configurations of the BZT SLs. The
$Amm2$ phase is found to be the ground state, which is used as the
reference energy. The total energies obtained in MC simulations are
decomposed to obtain the energies due to the dipole-dipole interaction
and the short-range interaction \cite{Zhong1995}. \label{tab:energy_comparision}}

\end{table*}
In Tab. \ref{tab:energy_comparision} we compare the energies of different
phases that have been described above. Among all the structures we
investigated, the $Amm2$ phase {[}Fig. \ref{fig:different_phase}(b){]}
has the lowest energy. On the other hand, the $P4mm$ phase {[}Fig.
\ref{fig:different_phase}(a){]}, which has dipoles along the $\left[001\right]$
direction has a much higher energy, which is $88.22$ (LDA) and $179.79$
(PBE) meV / 10 atoms higher than the $Amm2$ phase. It is this high
energy that prevents the development of polarization along the out-of-plane
polarization in the BZT SLs. Interestingly, the other three dipole
configurations (the $Pma2$, $Cmcm$ and $Pmc2_{1}$ phases) are all
extremely close in energy to that of the groundstate $Amm2$ phase
and are consistent with Ref. {[}\onlinecite{Lebedev2013}{]}. Similar
results provided MC simulations are given in the fourth column of
Tab. \ref{tab:energy_comparision}, where $Amm2$ is also found to
be the ground state. Moreover, given Fig. \ref{fig:different_phase}(e)
has a low energy, it is reasonable to say, if the periodicity along
$\left[100\right]$ is larger (i.e., the condensed phonon has a wavevector
of $X/3=\left(1/6,0,0\right)$, $X/4=\left(1/8,0,0\right)$ ... in
the Brillouin zone), the energy difference could be even smaller as
the portion of the interface energy (interface of two 90$^{\circ}$
domains) decreases. The closeness in energy of all these different
phases to the $Amm2$ shows that metastable states of almost degenerate
energy can exist in BZT SLs, which explains why our MC results show
complex dipole patterns {[}i.e., our earlier observations (i)--(iv){]}
that even include vortices and anti-vortices at 10 K. Therefore maintaining
a certain orientation of the polarization, which may be necessary
in certain application, in a given BTO layer is difficult at room
temperature, although it is shown the energy barrier between different
polarization orientations could be high \cite{Lebedev2013}, because
the various dipole patterns shown in Figs. \ref{fig:The-dipole-configuration}(c)--\ref{fig:The-dipole-configuration}(e)
and many others that correspond to phonons with small wavevectors
can bridge the transformation between different polarization. Further
investigation, for instance, on BZT SLs under mechanical strain, may
be necessary for the purpose of maintaining a certain polarization
in a given layer.

Since elastic energy due to lattice distortion can have a profound
effect on polarization \cite{Pertsev2007}, we decompose the total
energy obtained in MC simulations into its constitutional parts (similar
to what is done in a previous work \cite{Jiang2014}) and find that,
for the $Amm2$ phase {[}Fig. \ref{fig:different_phase}(b){]} at
5 K, the short-range interaction \cite{ShortRangeInteraction} is
195.86 meV/10 atoms and comparable in magnitude to that of the dipole-dipole
interaction (--370.17 meV/10 atoms). This result indicates that the
energy arising from both intra-plane and inter-plane couplings of
distorted lattice cells is an important part of the total energy.
Table \ref{tab:energy_comparision} also shows energies due to the
dipole-dipole interaction and the short-range interaction for various
dipole configurations. The tiny difference in the short-range interactions
between the $Amm2$, $Pma2$, and $Cmcm$ phases, which have very
different polarization in adjacent Ti layers, indicates that the elastic
couplings between Ti layers are very small. On the other hand, the
relatively large values of the $Pmc2_{1}$ phase indicate that intra-layer
interactions are stronger than that of the inter-layer. This situation
arises because the small lattice distortion in Zr layers (indicated
by the small dipoles associated with Zr sites) make them effective
buffer layers to reduce couplings between Ti layers, which is the
main cause of independent electric dipole sheets in BZT SLs. 

\begin{figure}
\includegraphics[width=8cm]{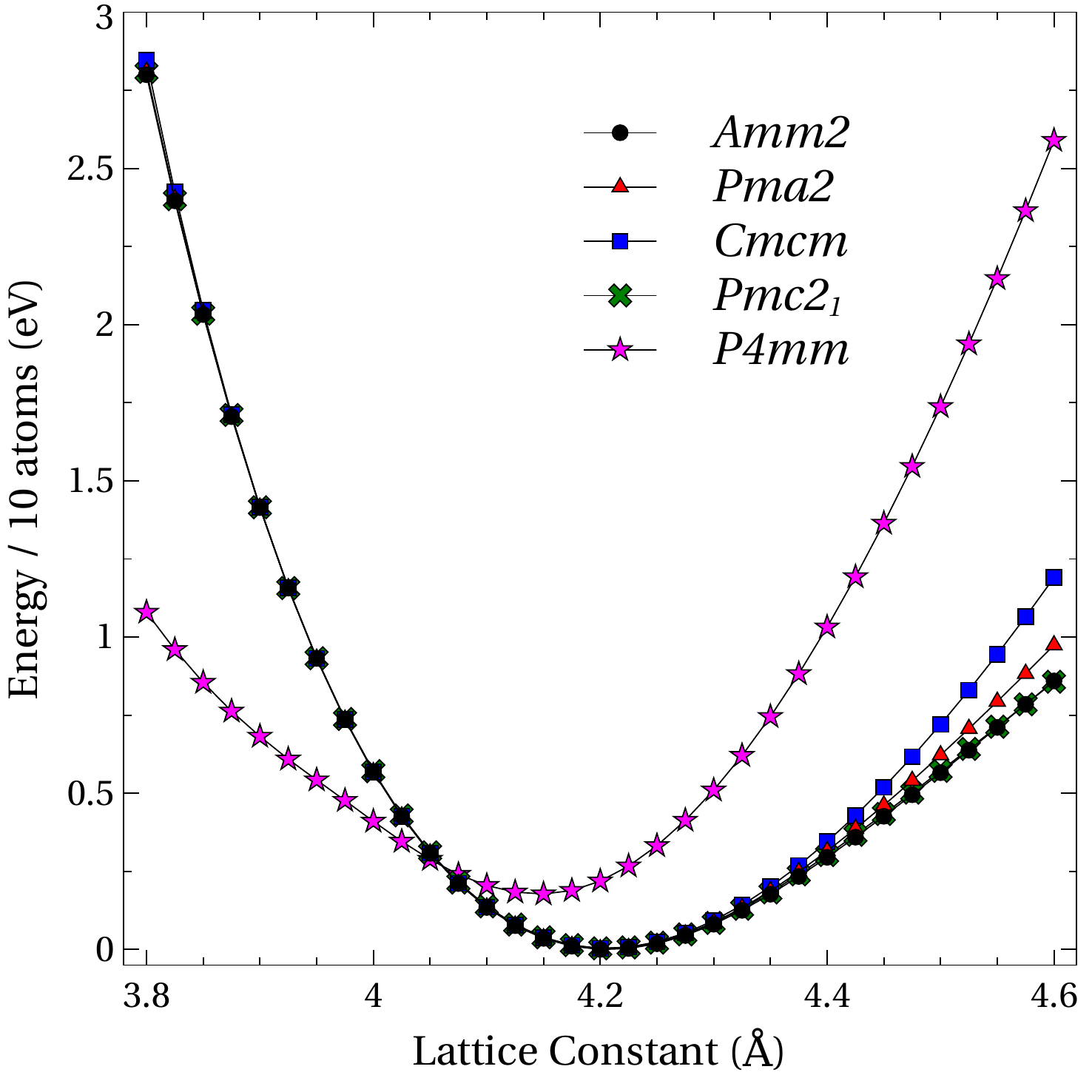}

\protect\caption{(Color online) Energy of various phases as a function of the in-plane
lattice parameter in epitaxial $(001)$ BZT SL. \label{fig:Strain}}
\end{figure}

Finally, since epitaxial strain can strongly shape polarization in
thin films \cite{Pertsev1998,Pertsev2003,Pertsev2007}, here we also
investigate how the energy of the various phases of BZT SLs change
with epitaxial strain. Figure \ref{fig:Strain} displays the total
energy of various phases as a function of the in-plane lattice constant
in films made of BZT SL. We numerically investigated how the epitaxial
strain can change dipole configurations in BZT SLs. For instance,
at a compressive strain ($\lesssim-4\%$), the $P4mm$ phase has a
lower energy than the $Amm2$ phase, thus become energetically favorable.
In contrast, the energy difference between the $Pma2$ (or $Cmcm$)
phase and the $Amm2$ phases increases quickly with a tensile strain,
and become larger than a hundred meV / 10 atoms at $\sim9\%$, thus
the $Pma2$ (or $Cmcm$) phase will likely disappear in the BZT SL
in favor of the $Amm2$ phase. Interestingly, the $Amm2$ and $Pmc2_{1}$
phases have the same energy for the all strain range (compressive
and tensile strains).

\subsection{Electrostriction}

\begin{figure}
\includegraphics[height=10cm]{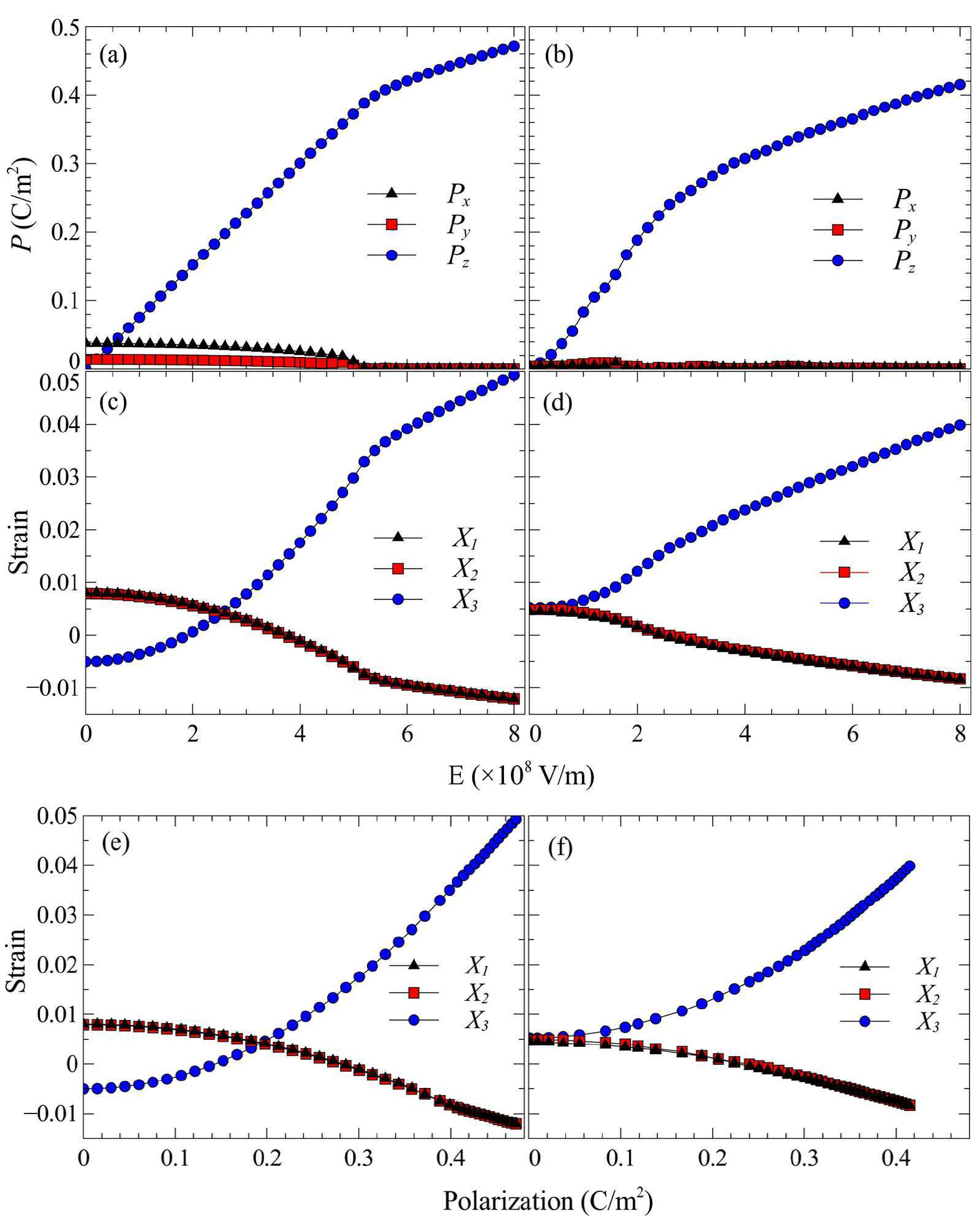}

\protect\caption{(Color online) Polarization (top panel) and strain (middle panel)
as a function of an electric field, strain as a function of polarization
(bottom panel) for the ordered (left column) and disordered BZT (right
column) at 10 K. The electric field is applied along the $[001]$
direction. We note the strain is calculated with reference to the
lattice constant $a_{0}=4.047\thinspace\textrm{\AA}$, which is a
starting parameter in MC simulations. \label{fig:Polarization-Strain-vs-E} }

\end{figure}

Having investigated the phase structures and dipole configurations
of the BZT SLs, we now use MC simulations to check if such a system
offers any improved properties over the disordered system. In particular,
we examine the polarization and strain variation versus a dc electric
field $E$ applied along the $\left[001\right]$ direction, which
provides information regarding static dielectric and electrostrictive
properties {[}Figs. \ref{fig:Polarization-Strain-vs-E}(a)--\ref{fig:Polarization-Strain-vs-E}(d){]}.
For (001)-ordered BZT, the polarization $P_{z}$ is linear with $E$
below $E=5.2\times10^{8}$ V/m and continues to be linear with $E$
above, but with a smaller slope (i.e., a smaller dielectric constant).
Interestingly, Fig. \ref{fig:Polarization-Strain-vs-E}(c) shows that
the strain versus $E$ curve is perfectly quadratic for $E<5.2\times10^{8}$
V/m, which implies a strong electrostrictive effect (but no piezoelectric).
As a comparison, we also simulated a disordered BZT, which are shown
in Figs. \ref{fig:Polarization-Strain-vs-E}(b) and \ref{fig:Polarization-Strain-vs-E}(d).
As we can see, for the dielectric constant, the first linear region
is smaller (below $E=2.5\times10^{8}$ V/m instead of $5.2\times10^{8}$
V/m). The values of the static relative permittivity in the linear
region for ordered and disordered systems are $84.7$ and $101.6$,\textbf{
}respectively. More significantly, the electrostrictive strain is
much smaller. Figures \ref{fig:Polarization-Strain-vs-E}(e) and \ref{fig:Polarization-Strain-vs-E}(f)
show the variation of homogeneous strain with respect to the polarization
for the ordered and disordered systems. To quantify the electrostriction
effect, we employ the expression $S_{ij}=Q_{ijkl}P_{k}P_{l}$ to fit
the curves in Figs. \ref{fig:Polarization-Strain-vs-E}(e) and \ref{fig:Polarization-Strain-vs-E}(f),
where $S_{ij}$ represents strain tensors and $Q_{ijkl}$ are electrostrictive
coefficients. It is found that for ordered BZT $Q_{11}=0.25$\,m$^{4}$/C$^{2}$
and $Q_{12}=0.10$\,m$^{4}$/C$^{2}$, meanwhile for the disordered
BZT $Q_{11}=0.19$\,m$^{4}$/C$^{2}$ and $Q_{12}=0.074$\,m$^{4}$/C$^{2}$.
The large theoretical values of electrostriction coefficients suggest
that BZT is worth further experimental investigation \cite{BTO-Q11}.
We also note the range of quadratic electrostriction coefficient in
disordered system is much smaller than the ordered BZT SLs. 

Moreover, we also obtained the polarization response to the electric
field (applied along the $\left[001\right]$ direction) at a compressive
strain of $-5\%$ (where $P4mm$ phase is used since it is the ground
state) and a tensile strain of $+9\%$ (where the $Amm2$ phase is
used ). We found that the the strain is linearly dependent on the
polarization $P_{z}$, resulting in pizeoelectric effect at a compressive
strain of $-5\%$. On the other hand, at the tensile strain of $+9\%$,
we find $Q_{11}=0.2$\,m$^{4}$/C$^{2}$, which is slightly smaller
than that of the BZT SLs without any in-plane strain constraint.

\section{Summary}

In summary, we have studied the dipole configurations and some basic
properties of BZT SLs, and shown that, due to the existence of metastable
states, complex dipole configurations and domains of different dipole
orientations may exist. It is worth noting that complex dipole patterns
are observed in BaTiO$_{3}$ quantum dots and wires \cite{Fu2003},
and in compositionally modulated (Ba,Sr)TiO$_{3}$ where geometric
frustration were observed \cite{Narayani2011}. Such energy degeneracy
and the resulting frustration of dipole configurations are mostly
known in magnetic systems, where the Ising model had been employed
for theoretical investigation \cite{Steele2011,Morgenstern1981,Gawlinski1985,Moessner2001,Iglesias2002,Jagla2004},
while much less known in electric dipole systems. With our findings
here, it can be concluded that, due to energy degeneracy, frustration-like
dipole configurations could exist in 0D, 1D, 2D, and 3D electric dipole
systems. 

In addition, it is rather interesting that such a conventional system
\cite{Tsurumi2002,Choudhury2008,Marssi2010} effectively form 2D electric
dipole sheets \cite{Lebedev2013}, possessing frustration-like states,
and strongly resembling magnetic sheets. We believe this special system
deserves more attention: (i) There are many possibilities to grow
SLs with BTO and BZO, for instance, along the $\left[110\right]$
and the $\left[111\right]$ directions (the Ti layers will have a
distance of $2a_{\textrm{lat}}/\sqrt{2}$ and $2\sqrt{3}a_{\textrm{lat}}/3$,
respectively, where $a_{\textrm{lat}}$ is the distance between a
Ti site and its nearest Zr site). It will be interesting to know if
such tuning of the distance between Ti ions will affect properties
of BZT or induce phenomena that are unknown in disordered BZT. In
particular, one may wonder if the relaxor properties can still exist
in ordered systems ; (ii) A systematical investigation on the effects
of epitaxial strain may reveal other novel phenomena as the misfit
strain is tuned. For instance, the $Pmc2_{1}$ could be the ground
state as it happens in PbTiO$_{3}$ and BiFeO$_{3}$ \cite{Yang2012,Yangyurong2014};
(iii) Complex dipole configurations (e.g., 2D vortex structures) in
the BTO/BZO 2D electric dipole system need to be experimentally investigated.
We note, recently the field-induced domain distortion in the PbTiO$_{3}$/SrTiO$_{3}$
SLs had been experimentally demonstrated \cite{QZhang2013}, showing
the feasibility of such investigations. Therefore by investigating
such a 2D system, we hope it expands our understanding of chemically
ordered BZT and will encourage other studies on this interesting system.
\\

\begin{acknowledgments}
Discussions with Drs. L. Bellaiche, S. Prosandeev, A. Bokov, J. Hlinka,
J. Petzelt, Z.-G. Gui, and X.-J. Lou are greatly acknowledged. This
work is financially supported by the National Natural Science Foundation
of China (NSFC) grant 51390472 and NSF grant DMR-1066158. F.L. acknowledges
the NSFC grant 51102193. Z.J. acknowledges the support from the 111
Project (B14040). We thank Shanghai Supercomputer Center for providing
resources for DFT computations. Some computations were also made possible
thanks to the MRI grant 0722625, MRI-R2 grant 0959124, and CI-TRAIN
grant 0918970 from NSF.\end{acknowledgments}


\begin{thebibliography}{10}
\bibitem{BZT-lattice-constant} The experimental lattice constant
of BaZrO$_{3}$ (BaTiO$_{3}$) is 4.19 (4.01)\,\AA{} and therefore
the misfit strain is estimated to be $\sim4\%$ considering the fact
that the lattice constant of the SL is between that of BaTiO$_{3}$
and BaZrO$_{3}$.
\end{thebibliography}
\end{document}